\documentstyle[aps,amstex,amssymb,preprint,epsf]{revtex}
\tightenlines
\begin{document}                                                       

\draft 

                                            
\title {Classical trajectories and quantum tunneling}

\author{Boris Ivlev} 

\address
{Department of Physics and Astronomy\\
University of South Carolina, Columbia, SC 29208\\
and\\
Instituto de F\'{\i}sica, Universidad Aut\'onoma de San Luis Potos\'{\i}\\
San Luis Potos\'{\i}, S. L. P. 78000 Mexico}

\maketitle

\begin{abstract}
The problem of inter-band tunneling in a semiconductor (Zener breakdown) in a nonstationary and homogeneous electric field 
is solved exactly. Using the exact analytical solution, the approximation based on classical trajectories is studied. A new 
mechanism of enhanced tunneling through static non-one-dimensional barriers is proposed in addition to well known normal 
tunneling solely described by a trajectory in imaginary time. Under certain conditions on the barrier shape and the particle 
energy, the probability of enhanced tunneling is not exponentially small even for non-transparent barriers, in contrast to 
the case of normal tunneling.  

\end{abstract} \vskip 1.0cm
   
 
\narrowtext
\section{INTRODUCTION}
\label{sec:intro}
A control of processes of quantum tunneling through potential barriers by external signals is a part of the field called 
quantum control which is actively developed now, see, for example, Ref.~\cite{RABITZ} and references therein. Excitation of
molecules, when one should excite only particular chemical bonds \cite{SHI,JUDSON,KOHLER}, formation of programmable atomic
wave packets \cite{SCHUMACHER}, a control of electron states in heterostructures \cite{KRAUSE,BERMAN}, and a control of 
photocurrent in semiconductors \cite{ATANASOV}, are typical examples of control by laser pulses. A control of quantum 
tunneling through potential barriers is also a matter of interest, since tunneling is a part of many processes in nature. 
The computation of probability for a classically forbidden region has a certain peculiarity from the mathematical stand 
point: there necessarily arises here the concept of motion in imaginary time or along a complex trajectory 
\cite{LANDAU,POKR,COLEMAN}. The famous semiclassical approach of Wentzel, Kramers, and Brillouin (WKB) \cite{LANDAU} for 
tunneling probability can be easily (in the case of a static potential) reformulated in terms of classical trajectories in 
complex time as a simple change of variables. The method of complex trajectories can also be applicable to a nonstationary 
case \cite{KELDYSH,PERELOMOV} which is not trivial. The method has been further developed in papers 
\cite{MELN1,MELN2,MELN3,MELN4,MELN5}, when singularities of the trajectories in the complex plane were accounted for an 
arbitrary potential barrier (see also \cite{MILLER}). Recent achievements in the semiclassical theory are present in 
Refs.\cite{KESHA,DEFENDI,MAITRA,ANKERHOLD,CUNIBERTI}.

Let us focus on the main aspects of tunneling under nonstationary conditions. When the electric field ${\cal E}\cos\Omega t$
acts on a tunneling particle of the initial energy $E$, it can absorb the quantum $\hbar\Omega$ (with the probability
proportional to the small parameter ${\cal E}^{2}$) and tunnel after that in a more transparent part of the barrier with
the higher energy $E+\hbar\Omega$. The pay in the absorption probability may be compensated by the probability gain in 
tunneling. In this case the system tends to absorb further quanta to increase the total probability of passing the barrier.
This mechanism of barrier penetration is called photon-assisted tunneling. If $\hbar\Omega$ is not big, the process of 
tunneling, with the simultaneous multi quanta absorption, can be described in a semiclassical way by the method of 
classical trajectories in the complex time \cite{MELN1,MELN2,MELN3,MELN4,MELN5}. When a tunneling particle of the energy 
$E$ is acted by a short-time pulse, the tunneling probability is associated with the particle density carrying away in
the outgoing wave packet. The particle energy after escape is $E+\delta E$, where the energy gain $\delta E=N\hbar\omega$, 
should be extremized with respect to the number of absorbed quanta $N$ and the energy $\hbar\omega$ of each quantum 
\cite{IVLEV1,IVLEV2}. This mechanism relates to semiclassical method for nonstationary potentials. 

The semiclassical method in quantum mechanics is very elegant and constructive. On the other hand, this method is based on 
the delicate mathematical issue called Stokes phenomenon when a solution is not expected to appear but it appears
\cite{HEADING}. This makes a use of the semiclassical method to be not trivial even in the case of a static potential. In
this situation the role of exactly solvable problems in quantum mechanics is very important. For example, in the problem
of reflection of a particle from certain static potentials one can follow in details a formation of a semiclassical 
solution from the exact one \cite{LANDAU}. Is it possible to find an exactly solvable problem for a nonstationary potential 
(excepting not interesting parabolic one, where the effect is trivial \cite{MELN4,MELN5}) to see how the exact solution is 
reduced to complex classical trajectories under semiclassical conditions? An exactly solvable problems would be extremely 
useful since the method of classical trajectories in nonstationary problems is still challenged and an exact analytical 
solution would place classical trajectories in the rank of mathematical theorem. 

Such exactly solvable nonstationary problem exists. This is an inter-band tunneling (Zener breakdown) in a semiconductor 
\cite{ZIMAN} in a nonstationary electric field which is a constant in space. The latter condition makes the problem to be
exactly solvable since in the momentum representation the Schr\"{o}dinger equation is reduced to the first order which can
be solved by the method of characteristics. In this paper it is studied how the exact solution turns over into one 
corresponding to classical trajectories and under which conditions this is possible. The goal of the paper is not to 
investigate Zener effect in real semiconductors but to use this situation to mathematically justify the method of classical 
trajectories.

Another issue of this paper is that a new mechanism of {\it enhanced} tunneling through static non-one-dimensional barriers 
is proposed in addition to well known {\it normal} tunneling solely described by a trajectory in imaginary time. As shown in 
the paper, under certain conditions on the barrier shape and the particle energy, the probability of enhanced tunneling is not 
exponentially small even for non-transparent barriers, in contrast to the case of normal tunneling.
\section{PHOTON-ASSISTED TUNNELING}
\label{sec:ph-as}
A penetration of a particle through a potential barrier is forbidden in classical mechanics. Only due to quantum effects 
the probability of passing across a barrier becomes finite and it can be calculated on the basis of WKB approach, which is 
also called the semiclassical theory. The transition probability through the barrier, shown in Fig.~\ref{fig1}, is
\begin{equation} 
\label{1}
W\sim\exp\left[-A_{0}(E)\right]
\end{equation}
where
\begin{equation}
\label{2}
A_{0}(E)=\frac{2}{\hbar}\int dx\sqrt{2m\left[V(x)-E\right]}
\end{equation}
is the classical action measured in units of $\hbar$. The integration goes under the barrier between two classical turning 
points where $V(x)=E$. One can use the general estimate $A_{0}\sim V/\hbar\omega$, where $V$ is the barrier height and $\omega$ is
the frequency of classical oscillations in the potential well. A semiclassical barrier relates to a big value 
$V/\hbar\omega\gg 1$. 

What happens when the static potential barrier $V(x)$ is acted by a weak nonstationary electric field ${\cal E}(t)$? In this
case there are two possibilities for barrier penetration: (i) the conventional tunneling, which is not affected by 
${\cal E}(t)$, shown by the dashed line in Fig.~\ref{fig2}(a), and (ii) an absorption of the quantum $\hbar\Omega$ of the field
${\cal E}(t)$ and subsequent tunneling with the new energy $E+\hbar\Omega$. The latter process is called photon-assisted 
tunneling. The total probability of penetration across the barrier can be schematically written as a sum of two 
probabilities
\begin{equation}
\label{3}
W\sim\exp\left(-\frac{V}{\hbar\omega}\right)+
\left(\frac{a\hspace{0.05cm}{\cal E}_{\Omega}}{\hbar}\right)^{2}\exp\left(-\frac{V-\hbar\Omega}{\hbar\omega}\right)
\end{equation}
where ${\cal E}_{\Omega}$ is the Fourier component of the field ${\cal E}(t)$ and the length $a$ is a typical barrier 
extension in space. The second term in Eq.~(\ref{3}) relates to photon-assisted tunneling and it is a product of 
probabilities of two quantum mechanical processes: absorption of the quantum $\hbar\Omega$ and tunneling through the 
reduced barrier $V-\hbar\Omega$. Since in quantum mechanics one should add amplitudes but not probabilities, Eq.~(\ref{3}) 
is rather schematic and serves for general illustration only. For example, the accurate perturbation theory starts with a 
linear ${\cal E}$-term. When the frequency is high $\Omega >\omega$, the second dominates at sufficiently small nonstationary 
field $(a{\cal E}_{\Omega}/\hbar)^{2}>\exp(-\Omega/\omega)$. This is a feature of tunneling processes since, normally, a 
nonstationary field dominates at bigger amplitudes $(a{\cal E}_{\Omega}/\hbar)^{2}>1$. When the second term in Eq.~(\ref{3}) 
exceeds the first one, further orders of perturbation theory should be accounted which correspond to the multiple absorption, 
shown in Fig.~\ref{2}(b). 

Let us specify a shape of a field pulse in the form
\begin{equation}
\label{4}
{\cal E}(t)=\frac{{\cal E}}{\left(1+t^{2}/\theta^{2}\right)^{3}}
\end{equation}
with the Fourier component ${\cal E}_{\Omega}\sim{\cal E}\theta\exp(-\theta|\Omega|)$. In this case, in addition to the almost 
steady flux from the barrier, an outgoing wave packet (of the maximum amplitude $\sqrt{W}$) is created which carries away a certain 
particle density. The probability of the transition through the process of absorption of $N$ quanta and subsequent tunneling 
with the higher energy $E+N\hbar\Omega$, shown in Fig.~\ref{2}(b), can be estimated as
\begin{equation} 
\label{5}
w\sim\left(\frac{a{\cal E}_{\Omega}}{\hbar}\right)^{2N}\exp\left[-A_{0}(E+N\hbar\Omega)\right]=
\left(\frac{\pi}{\hbar}\hspace{0.05cm}\theta a{\cal E}\right)^{2N}\exp(-A)
\end{equation}
where
\begin{equation}
\label{6}
A=\frac{2\theta}{\hbar}\hspace{0.1cm}\delta E + A_{0}(E+\delta E)
\end{equation}
Here the total energy transfer $\delta E = N\hbar\Omega$ is introduced ($\Omega > 0$). The maximum squared amplitude $W$  of 
the outgoing packet is associated with a maximum value of $w$ and is determined by the extreme $\delta E$ to get a minimum of 
$A$. This results in the condition $\partial A(E+\delta E)/\partial\delta E = 0$. An existence of such a minimum is possible 
if growing up of small $\delta E$ reduces $A$, that is, under the condition $2\theta <\hbar\mid\partial A_{0}(E)\partial E\mid$ 
of sufficiently short pulses. In other words, sufficiently short and not very small pulses (however, still much smaller than 
the static barrier field) strongly enhance tunneling by photon assistance. 

In principal, this type of treatment of nonstationary tunneling is reasonable since a maximum of the outgoing packet should
relate to some extreme condition. Nevertheless, the approach (\ref{5}) is schematic since a quantum interference between 
absorption and tunneling is neglected. What is a ``scientific'' way to account precisely a nonstationary field in tunneling? 
We go towards this in \ref{sec:traj}.
\section{TRAJECTORIES IN IMAGINARY TIME}
\label{sec:traj}
\begin{figure}[p]
\begin{center}
\vspace{1.5cm}
\leavevmode
\epsfxsize=\hsize
\epsfxsize=13cm
\epsfbox{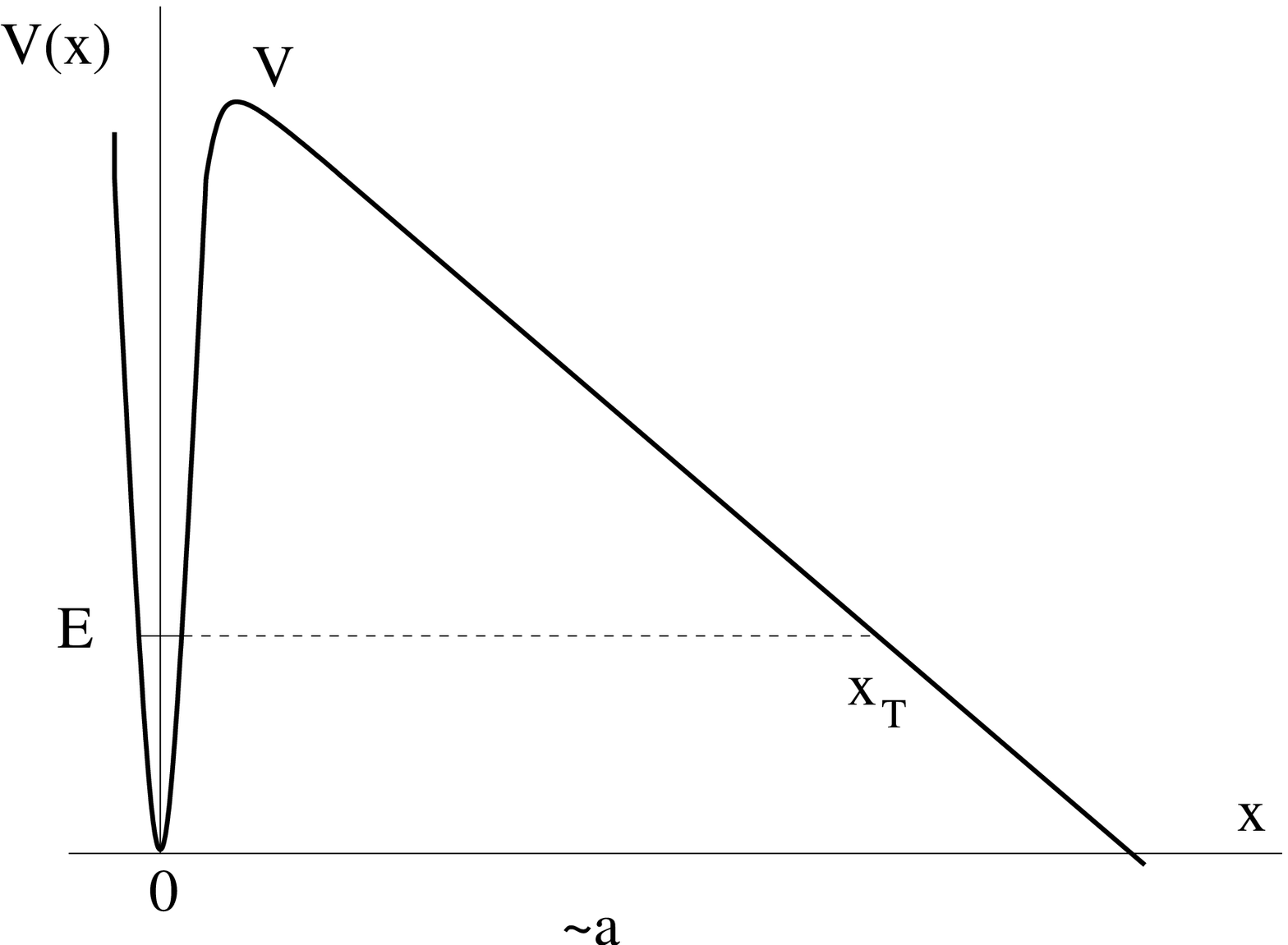}
\vspace{2cm}
\caption{The path of tunneling is denoted by the dashed line. $x_{T}$ is the classical turning point. 
$E$ is the energy of the metastable state, $V$ is the barrier height, and $a$ is the typical 
potential length.}
\label{fig1}
\end{center}
\end{figure}

\begin{figure}[p]
\begin{center}
\vspace{1.5cm}
\leavevmode
\epsfxsize=\hsize
\epsfxsize=12cm
\epsfbox{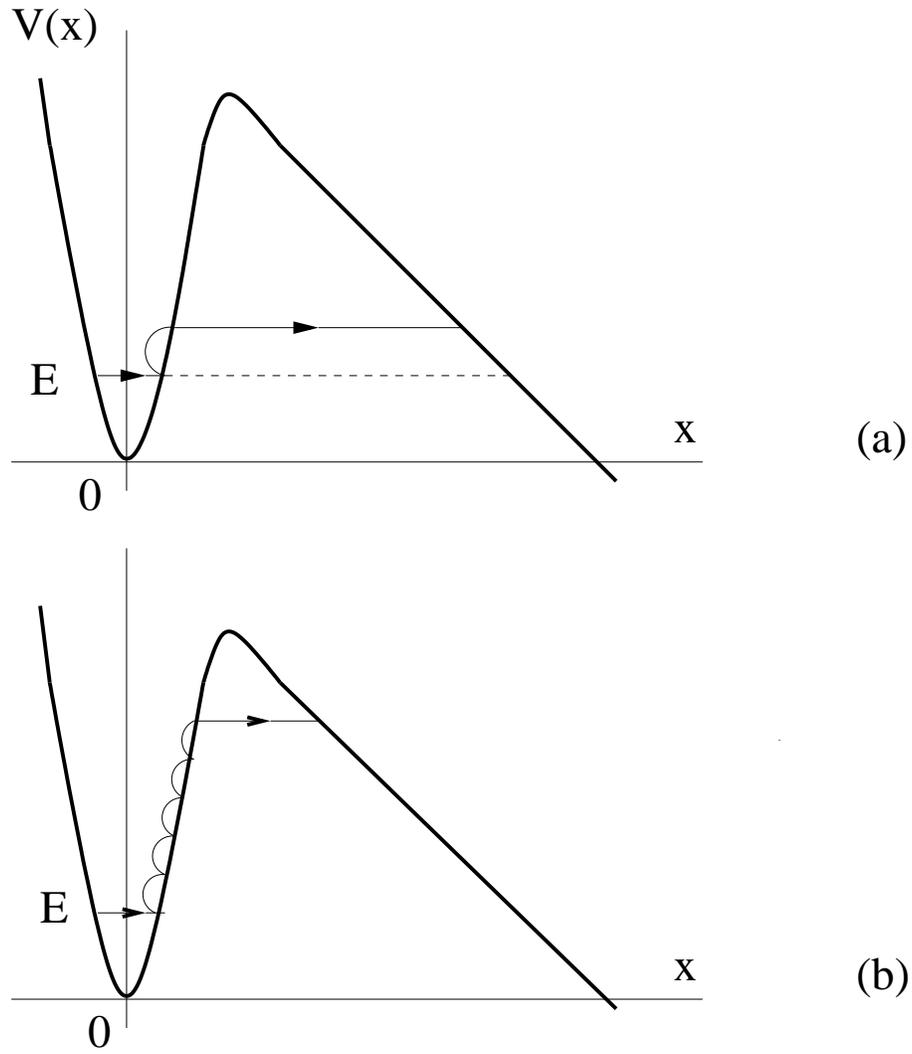}
\vspace{2cm}
\caption{(a) The particle can absorb a quantum and tunnel in a more transparent part of the barrier with the energy 
$E+\hbar\Omega$; (b) the process of the multi-quanta absorption with the subsequent tunneling.}
\label{fig2}
\end{center}
\end{figure}

According to Feynman \cite{FEYNMAN}, when the phase of a wave function is big, it can be expressed through classical 
trajectories of the particle. But in our case there are no conventional trajectories since a classical motion is forbidden
under a barrier. Suppose a classical particle to move in the region to the right of the classical turning point $x_{T}$ in 
Fig.~\ref{fig1} and to reach this point at $t=0$. Then, close to the point $x_{T}$, $x(t)=x_{T}+ct^{2}$ ($c>0$) and there is no 
a barrier penetration as at all times $x(t)>x_{T}$. Nevertheless, if $t$ is formally imaginary, $t=i\tau$, the penetration 
becomes possible since $x(i\tau)=x_{T}-c\tau^{2}$ is less then $x_{T}$. Therefore, one can use classical trajectories in 
imaginary time to apply Feynman's method to tunneling. In the absence of a nonstationary field a classical trajectory 
satisfies Newton's equation in imaginary time
\begin{equation}
\label{7}
m\hspace{0.1cm}\frac{\partial^{2}x}{\partial\tau^{2}}=\frac{\partial V(x)}{\partial x}
\end{equation}
where $V(x)$ is the static barrier in Fig.~\ref{fig1}. The classical turning point $x_{T}$ in Fig.~\ref{fig1} is reached at
$\tau =0$ with the initial condition $\partial x/\partial\tau =0$. The classical trajectory $x=x(\tau)$ can be considered
as a change of variables $x\rightarrow\tau$ in the WKB exponent (\ref{2}) when it becomes of the form
\begin{equation}
\label{8}
A_{0}=\frac{2}{\hbar}\int d\tau\left[\frac{m}{2}\left(\frac{\partial x}{\partial\tau}\right)^{2}+V(x)-E\right]
\end{equation}
So, in the absence of a non-stationary field, use of classical trajectories in imaginary time is obvious and it is simply
reduced to a change of variable. 

Is it possible to extend the method of classical trajectories in tunneling to a non-stationary case? 

\begin{figure}[p]
\begin{center}
\vspace{1.5cm}
\leavevmode
\epsfxsize=\hsize
\epsfxsize=10cm
\epsfbox{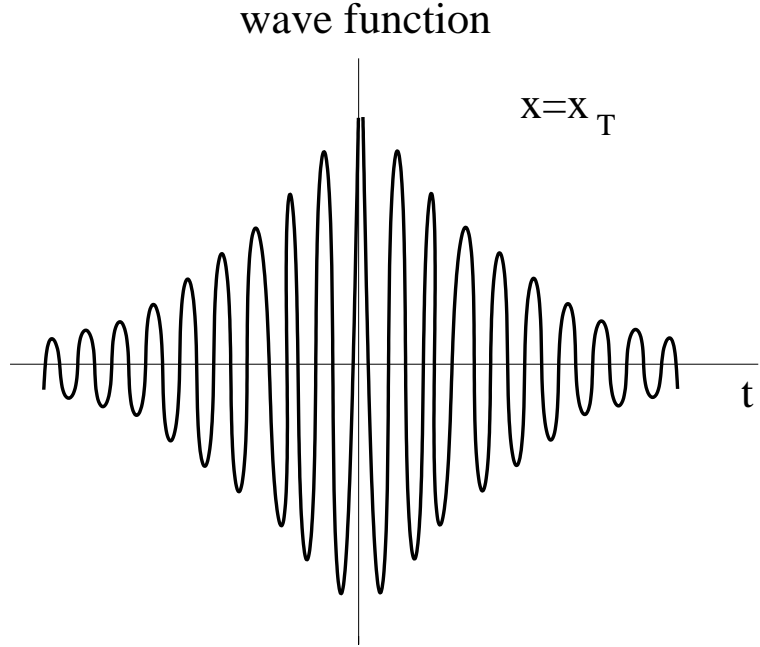}
\vspace{1cm}
\caption{The outgoing wave function at the turning point $x_{T}$ has a maximum amplitude under the 
influence of the external pulse.}
\label{fig3}
\end{center}
\end{figure}

Suppose some pulse of an external field, for example (\ref{4}), acts on a tunneling particle. At the point $x_{T}$, besides 
an almost constant background, there is a wave packet of the outgoing particles shown in Fig.~\ref{fig3}. If the method of 
classical trajectories is applicable, the maximum of the squared amplitude of the outgoing packet 
\begin{equation}
\label{9}
W\sim\exp(-A)
\end{equation}
should correspond, with the exponential accuracy, to the classical action 
\begin{equation}
\label{10}
A=\frac{2}{\hbar}\int^{\tau_{0}}_{0} d\tau\left[\frac{m}{2}\left(\frac{\partial x}{\partial\tau}\right)^{2}+V(x)-
x{\cal E}(i\tau)-E\right]
\end{equation}
since a classical trajectory relates to an extreme value of the action. The classical trajectory should satisfy Newton's 
equation
\begin{equation}
\label{11}
m\hspace{0.1cm}\frac{\partial^{2}x}{\partial\tau^{2}}-\frac{\partial V(x)}{\partial x}=-{\cal E}(i\tau)
\end{equation}
with the conditions
\begin{equation}
\label{12}
\frac{\partial x}{\partial\tau}\bigg |_{\tau =0}=0;\hspace{1cm}x(i\tau_{0})=0
\end{equation}
We suppose the potential well to be narrow and localized close to $x=0$ as in Fig.~\ref{fig1}. The under-barrier 
``time'' $\tau_{0}$ is expressed through the particle energy $E$ in the well
\begin{equation}
\label{13}
E=\frac{m}{2}\left(\frac{\partial x}{\partial\tau}\right)^{2}_{\tau_{0}}+V(0)
\end{equation}
which is weakly violated by a small non-stationary field. Certain semiclassical conditions (not extremely small 
and sufficiently slow varying ${\cal E}(t)$) should be fulfilled. 

Eqs.~(\ref{9}-\ref{13}) constitute the method of classical trajectories in tunneling under non-stationary 
conditions. This method relates to calculation solely of the maximum in time $W_{max}$ of the probability $W(t)$ 
of outgoing particles. In contrast to a static barrier, the method of classical trajectories is not trivial in 
application to a nonstationary case. This can be seen from that the non-stationary field of the type 
${\cal E}\cos\Omega t$ in imaginary time becomes proportional to $\cosh\Omega\tau$ and may be very big. The same
relates to the field (\ref{4}) which is singular in imaginary time. This means, formally, that arbitrary weak 
amplitude ${\cal E}$ of a nonstationary field may produce a big effect. The arguments in \ref{sec:ph-as} also show 
an increase of an effective ${\cal E}$ in tunneling, but it is clear physically that an effect of an extremely small 
nonstationary field is negligible and the method of classical trajectories should not work in this case. As one can 
see, the method of classical trajectories is delicate and requires an accurate justification.

In the papers \cite{IVLEV1,IVLEV2} for the certain particular potential well and nonstationary pulses the wave 
function has been found in the form of expansion with respect to powers of $\hbar$
\begin{equation}
\label{14}
\psi (x,t)=\left[a(x,t)+\hbar b(x,t)+...\right]\exp\left[\frac{i}{\hbar}\hspace{0.1cm}S(x,t)\right]
\end{equation}
where $S$ is the classical action. The semiclassical formalism, developed in \cite{IVLEV1,IVLEV2}, confirms the 
method of classical trajectories and coincides with the general scheme (\ref{9}-\ref{13}). Nevertheless, it would 
be extremely instructive to find any exactly solvable case of tunneling under nonstationary conditions in order 
to see how the method of classical trajectories follows not from a semiclassical formalism but from an exact 
mathematical solution. In \ref{sec:zener} this program is completed.
\section{EXACT SOLUTION OF A TUNNELING PROBLEM UNDER NONSTATIONARY CONDITIONS}
\label{sec:zener}
\begin{figure}[p]
\begin{center}
\vspace{1cm}
\leavevmode
\epsfxsize=\hsize
\epsfxsize=8cm
\epsfbox{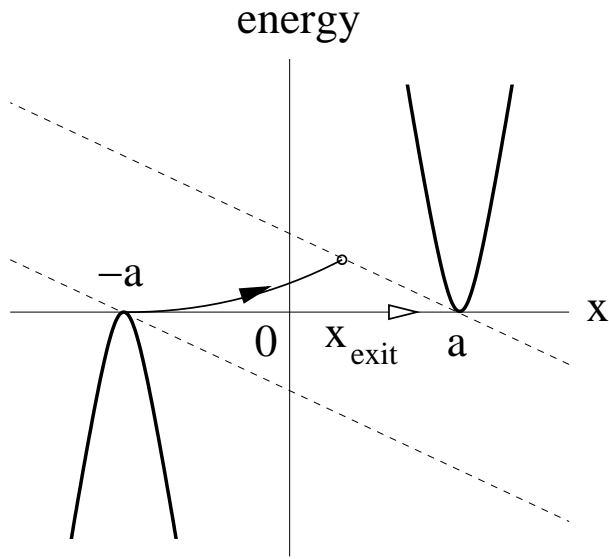}
\vspace{1cm}
\caption{An inter-band tunneling (Zener breakdown) in a semiconductor with the energy gap $\varepsilon_{g}$ 
in a static electric field ${\cal E}_{0}$, shown by the open arrow. The tunneling length is 
$2a=\varepsilon_{g}/{\cal E}_{0}$. With a nonstationary field the path inside the forbidden region is
shown by the full arrow.}
\label{fig4}
\end{center}
\end{figure}

Let us consider an one dimensional two band semiconductor in an external homogeneous electric field ${\cal E}_{0}+{\cal E}(t)$,
where ${\cal E}_{0}$ is the static component. The Schr\"{o}dinger equation has the form
\begin{equation}
\label{15}
i\hbar\hspace{0.1cm}\frac{\partial\psi}{\partial t}=
\frac{\varepsilon_{g}}{2}\hspace{0.1cm}\sqrt{1+\left(\frac{2c}{\varepsilon_{g}}\hspace{0.1cm}\hat{p}\right)^{2}}\hspace{0.2cm}
\psi -\left[{\cal E}_{0}+{\cal E}(t)\right]\psi
\end{equation}
where $\hat{p}=-i\hbar\partial/\partial x$, $\varepsilon_{g}$ is the energy gap, and $c$ is the velocity at big momentum. The 
square root in Eq.~(\ref{15}) may have two signs according to two energy bands. In a static electric field ${\cal E}_{0}$ 
the inter-band quantum tunneling is possible which is called Zener breakdown \cite{ZIMAN}. The incident flux of particles from 
left to the right in Fig.~\ref{fig4} is mainly reflected back from the tilted energy gap but a small fraction penetrates
the other band and goes to $+\infty$ with the probability \cite{ZIMAN}
\begin{equation}
\label{16}
W_{0}=\exp\left(-\frac{\pi\varepsilon_{g}t_{0}}{2\hbar}\right)
\end{equation}
where $t_{0}=a/c$ and $2a=\varepsilon_{g}/{\cal E}_{0}$ is the tunneling length in Fig.~\ref{fig4}. Eq.~(\ref{16}) holds
under the semiclassical condition
\begin{equation}
\label{16a}
\hbar\ll\varepsilon_{g}t_{0}
\end{equation}
The peculiarity of this problem is that it can be solved exactly with a nonstationary field by making the Fourier transformation 
with respect to $x$. In the new variables, momentum $q$ and time $t$, the Schr\"{o}dinger equation (\ref{15}) is of the first 
order with respect to derivatives $\partial /\partial q$ and $\partial /\partial t$ and can be solved by the method of 
characteristics. If to measure the coordinate in units of $a$ and time in units of $t_{0}$, the solution of Eq.~(\ref{15}) has 
the form
\begin{equation}
\label{17}
\psi(x,t)=\int^{\infty}_{-\infty}\frac{dq}{2\pi}\exp\left\{\frac{i\varepsilon_{g}t_{0}}{2\hbar}\left[qx-F_{t}(q)\right]\right\}
\end{equation}
where  
\begin{equation}
\label{18}
F_{t}(q)=\int^{q}_{-\infty}duf_{t}(u,q);\hspace{1cm}f_{t}(u,q)=\sqrt{1+\left[u+\int^{u-q}_{0}dsh(s+t)\right]^{2}}
\end{equation}
and  $h(t)={\cal E}(t)/{\cal E}_{0}$. Below we consider the case when the nonstationary field is much less compared to the
static value
\begin{equation}
\label{18a}
{\cal E}(t)\ll{\cal E}_{0}
\end{equation}
If ${\cal E}(t)$ is an analytical function of $t$, $F_{t}(q)$ is an analytical function of the complex argument $q$ and it has 
only two branch points at $q=\pm i$. According to this, Eqs.~(\ref{17}-\ref{18}) can be written through the complex paths 
\begin{equation}
\label{19}
\psi(x,t)=\int_{C}\frac{dq}{2\pi}\exp\left\{\frac{i\varepsilon_{g}t_{0}}{2\hbar}\left[qx-F_{t}(q)\right]\right\};\hspace{1cm}
F_{t}(q)=\int_{C{q}}duf_{t}(u,q)
\end{equation}
where the contour $C$ on the complex plane of $q$ and the contour $C_{q}$ on the complex plane of $u$ are shown in Fig.~\ref{5}.
Two cuts are denoted in Fig.~\ref{5}(a) by solid lines and the regular branch of the function $F_{t}(q)$ is chosen in order to 
get the limit
\begin{equation}
\label{20}
F_{t}(q)=-\int^{q}_{-\infty}du|f_{t}(u,q)|;\hspace{1cm}q\rightarrow -\infty
\end{equation}
The integral (\ref{20}) is divergent and has to be cut off at the lower limit by a big negative value which sets an
irrelevant constant phase shift. Under the semiclassical condition (\ref{16a}) the $q$-integration in Eq.~(\ref{19}) goes 
mainly in the vicinity of the saddle point(s) determined by the equation
\begin{equation}
\label{21}
x=\frac{\partial}{\partial q}\hspace{0.1cm}F_{t}(q)
\end{equation}
Suppose the nonstationary field ${\cal E}(t)\rightarrow 0$ at $t\rightarrow -\infty$ when the particle energy is zero 
and the particle goes to the tilted energy gap in Fig.~\ref{fig4} from the left. For $x$ to the left of the point $-a$ 
in Fig.~\ref{fig4} ($x<-1$ in the dimensionless units used) an influence of the nonstationary field under the condition 
(\ref{18a}) is small and two saddle points $q=\pm\sqrt{x^{2}-1}$ are shown in Fig.~\ref{fig6}(a). They correspond to the
incident and the reflected de Broglie waves. Since the tunneling probability is small, at $x<-1$ the wave function can be
shown to have the form
\begin{equation}
\label{22}
\psi(x,t)\sim\cos\left(\frac{\varepsilon_{g}t_{0}}{4\hbar}x^{2}-\frac{\pi}{4}\right)
\end{equation}
In the classically allowed regions, to the left and to the right of the tilted gap, the classical trajectory $x_{cl}(t)$               
(the equation, it satisfies, is written below) is weakly violated by the non-stationary field under the condition (\ref{18a})
\begin{equation}
\label{23}
x_{cl}(t)=
\begin{cases}
-\sqrt{1+t^{2}};&t<0\\
x_{exit}+\sqrt{1+t^{2}}-1;&t>0
\end{cases}
\end{equation}
excepting that $x_{exit}\neq 1$ now. The energy of the incident particle is zero and the energy after exit from the 
tilted gap conserves and equals $E_{exit}=(1-x_{exit})\varepsilon_{g}/2$.

\begin{figure}[p]
\begin{center}
\vspace{1.5cm}
\leavevmode
\epsfxsize=\hsize
\epsfxsize=15cm
\epsfbox{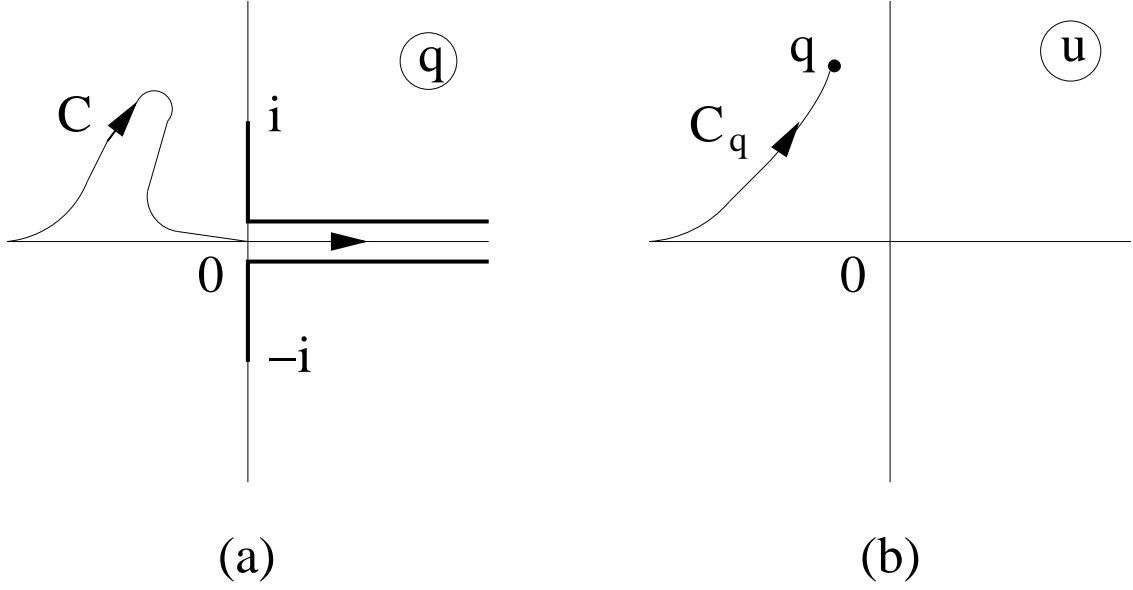}
\vspace{1cm}
\caption{(a) The cuts are shown by solid lines; (b) the contour $C_{q}$ terminates at the point $q$.}
\label{fig5}
\end{center}
\end{figure}

\begin{figure}[p]
\begin{center}
\vspace{1.5cm}
\leavevmode
\epsfxsize=\hsize
\epsfxsize=15cm
\epsfbox{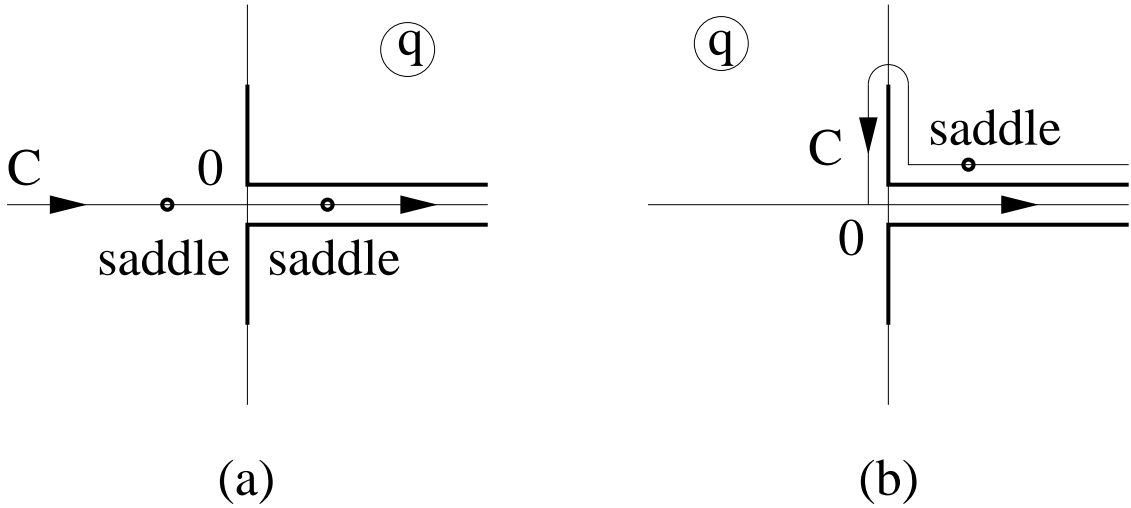}
\vspace{1cm}
\caption{(a) At $x<-a$ two saddle point are involved which correspond to the incident and the reflected de Broglie waves; 
(b) for positive $x$ the saddle point relates to a particle after tunneling.}
\label{fig6}
\end{center}
\end{figure}

For each $x$ and $t$ the saddle point condition (\ref{21}) determines a certain $q$. In the absence of a nonstationary 
field the amplitude of the outgoing wave is almost stationary. Under action of a pulse there is an additional outgoing 
wave packet at $t>0$ which goes to the right and keeps the constant maximum amplitude (if to neglect small quantum effects 
of smearing) at the classical trajectory (\ref{23}). We are interested to find this maximum amplitude of the outgoing packet
since it defines, with the exponential accuracy, a total number of particles passing the barrier. With $x=x_{cl}(t)$ the 
saddle point condition (\ref{21}) is satisfied by $q=t$ since $h(t)$ is small for real $t$. In order to reach the saddle 
point at positive $x$, one should deform the contour $C$ as shown in Fig.~\ref{fig6}(b). This could be done with no problems 
since $F_{t}(q)$ has only the branch point at $q=i$ in the upper half plane. The main contribution to the $q$-integration
over the bent contour $C$ comes from the saddle point, shown in Fig.~\ref{fig6}(b). The saddle point condition (\ref{21})
now reads
\begin{equation}
\label{24}
x_{exit}=1-2\int^{\tau_{0}}_{0}d\tau\hspace{0.1cm}\frac{\tau +
\int^{\tau}_{0}d\tau_{1}h(i\tau_{1})}{\sqrt{1-\left[\tau+\int^{\tau}_{0}d\tau_{1}h(i\tau_{1})\right]^{2}}}
\hspace{0.1cm}h(i\tau)
\end{equation}
The nonstationary field is supposed to be symmetric ${\cal E}(t)={\cal E}(-t)$, as the field (\ref{4}), so that 
${\cal E}(i\tau)$ is real. $\tau_{0}$ is a smallest positive root of the equation
\begin{equation}
\label{25}
\left[\tau_{0}+\int^{\tau_{0}}_{0}d\tau h(i\tau)\right]^{2}=1
\end{equation}
and can be called the tunneling time.
If to define the transition probability $W$ as a square of the ratio of the maximum amplitude of the outgoing 
packet and the amplitude of the incident wave, Eq.~(\ref{17}) gives with the exponential accuracy  
\begin{equation}
\label{26}
W\sim\exp\left(-A\right)
\end{equation}
where
\begin{equation}
\label{27}
A=\frac{2\varepsilon_{g}t_{0}}{\hbar}\int^{\tau_{0}}_{0}d\tau\sqrt{1-\left[\tau+\int^{\tau}_{0}d\tau_{1}
h(i\tau_{1})\right]^{2}}
\end{equation}
The action $A$ is collected due to the $q$-integration in the narrow vicinity of the right-hand-side saddle in 
Fig.~\ref{fig6}(b) where $q$ is real and positive. For such $q$ the contour $C_{q}$ in Fig.~\ref{fig5}(b) goes
around the branch point $u=i\tau_{0}$ and leads to the relation (\ref{27}). Without a non-stationary field 
Eqs.~(\ref{26}) and (\ref{27}) give the static result (\ref{16}). 

Using of the above method of saddle point means that the length $\delta q\sim\sqrt{\hbar/\varepsilon_{g}t_{0}}$ of the 
steepest descent integration should be sufficiently short compared to the typical scale of the problem 
($(\tau -\theta)/t_{0}$ in dimension units) which can be estimated from the relation $\int d\tau h(i\tau)\sim 1$. For 
the pulse (\ref{4}) the condition reads
\begin{equation}
\label{28}
\frac{\hbar}{\varepsilon_{g}t_{0}}\ll\left(\frac{\theta}{t_{0}}\right)^{3}\frac{{\cal E}}{{\cal E}_{0}} 
\end{equation}
In presence of a nonstationary field, the semiclassical condition (\ref{16a}) should be supplemented by the
additional semiclassical condition (\ref{28}). According to it, the nonstationary field ${\cal E}$ and its duration 
$\theta$ should not be too small. Note, that under the semiclassical condition (\ref{28}) the nonstationary amplitude
${\cal E}$ can be still smaller than the static field ${\cal E}_{0}$.

In Refs. \cite{MELN4,MELN5} the method of classical trajectories was developed for the present problem of 
inter-band tunneling in presence of a nonstationary field. According to this method, one has to minimize the classical 
action of a particle with the spectrum, as in (\ref{15},) and in a homogeneous electric field
\begin{equation}
\label{29}
S=\frac{\varepsilon_{g}t_{0}}{2}\int dt\left\{\sqrt{1-\left(\frac{\partial x}{\partial t}\right)^{2}}+
x\left[1+h(t)\right]\right\}
\end{equation}
to find a classical trajectory. It satisfies the equation
\begin{equation}
\label{30}
\frac{\partial x_{cl}(t)}{\partial t}=
\frac{t+\int^{t}_{0}dsh(s)}{\sqrt{1+\left[t+\int^{t}_{0}dsh(s) \right]^{2}}}
\end{equation}
The classical trajectory starts at $t=0$ at the point $x=x_{exit}$ in Fig.~\ref{fig4}. At the ``moment'' $i\tau_{0}$
it reaches some point inside the gap where two branches merge and, turning on the opposite side of the cut, it comes 
at $t=0$ in the point $x=-a$. Within the formalism, developed in Refs. \cite{MELN4,MELN5}, in order to obtain a tunneling 
probability, one should substitute the classical trajectory into the action (\ref{29}) and to integrate over the branch 
point $i\tau_{0}$. This procedure leads to the result (\ref{27}). 

For the nonstationary pulse (\ref{4}) under the condition (\ref{18a}), the tunneling time $\tau_{0}\simeq 1$ at 
$t_{0}<\theta$ and $\tau_{0}\simeq \theta /t_{0}$ at $\theta <t_{0}$. The action (\ref{27}) becomes of the form
\begin{equation}
\label{31}
A=\frac{\varepsilon_{g}t_{0}}{\hbar}
\begin{cases}
\pi/2\hspace{0.1cm};&t_{0}<\theta\\
\arcsin\left(\theta/t_{0}\right)+
\left(\theta/t_{0}\right)\sqrt{1-\left(\theta/t_{0}\right)^{2}}\hspace{0.1cm};&\theta <t_{0}
\end{cases}
\end{equation}
This expression weakly depends on ${\cal E}$ under the semiclassical condition (\ref{28}) since the role of the 
nonstationary field (\ref{4}), due to its singularity in imaginary time, is only to set the new tunneling time 
$\tau_{0}$. Obtained results for inter-band tunneling under nonstationary conditions may be interpreted as 
positive (positive energy transfer) photon-assisted tunneling shown in Fig.~\ref{fig4}. In this problem cannot be
such phenomenon as Euclidean resonance \cite{IVLEV2}, associated with a negative energy transfer.

Analogously one can consider the monochromatic ${\cal E}\cos\Omega t$ \cite{MELN4,MELN5} and the Gaussian 
${\cal E}\exp(-\Omega^{2} t^{2})$ nonstationary fields which become exponentially big in imaginary time and to establish
how much they can grow within the formalism of classical trajectories.

One can conclude now, that for inter-band tunneling in a nonstationary field under semiclassical conditions the method 
of classical trajectories follows from the exact analytical solution. The bridge between exact theory and classical
trajectories enables in problems of tunneling to treat a nonstationary field in complex time. 
\section{A NEW ENHANCED TUNNELING THROUGH STATIC BARRIERS}
\label{sec:static}
\begin{figure}[p]
\begin{center}
\vspace{1.5cm}
\leavevmode
\epsfxsize=\hsize
\epsfxsize=8cm
\epsfbox{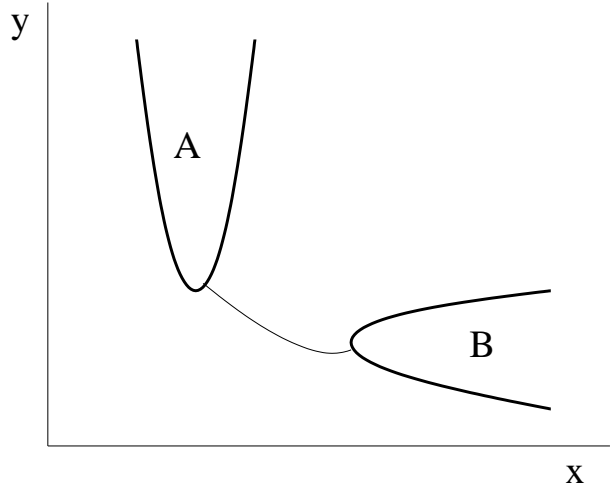}
\vspace{1cm}
\caption{Thick curves relate to the condition $V(x,y)=E_{0}$. Normal tunneling between two classically allowed region 
$A$ and $B$, where $V(x,y)<E_{0}$, is described by the classical trajectory in imaginary time (thin curve) which meets 
the borders of $A$ and $B$ normally.}
\label{fig7}
\end{center}
\end{figure}

Tunneling through one-dimensional static barrier is a well known problem described in almost every text book on quantum 
mechanics and the tunneling rate, with exponential accuracy, is given by WKB formulas (\ref{1}) and (\ref{2}). Tunneling 
through a two- or many-dimensional potential barrier was considered in a variety of publications
\cite{KAGAN,POKR,OKUN,STONE,COLEMAN,SCHMIDT,MELN6,OVCH}. The condition $V(\vec r)=E_{0}$, where $E_{0}$ is the total energy, 
determines certain surfaces in the coordinate space.  For a two-dimensional case it is shown in Fig.~\ref{fig7}. Tunneling 
probability from the classically allowed region $A$ to another such region $B$ is given by Eq.~(\ref{1}) where the exponent
\begin{equation}
\label{32}
A_{0}(A\rightarrow B)=\frac{2}{\hbar}\int^{\tau_{0}}_{0}d\tau\left[\frac{m}{2}\left(\frac{\partial\vec r}{\partial\tau}
\right)^{2}+V(\vec r)-E_{0}\right]
\end{equation}  
is expressed through the classical trajectory $\vec r(\tau)$ satisfying Newton's equation in imaginary time
\begin{equation}
\label{33}
m\hspace{0.1cm}\frac{\partial^{2}\vec r}{\partial\tau^{2}}=\frac{\partial V(\vec r)}{\partial\vec r}
\end{equation} 
The trajectory connects different surfaces, being perpendicular to them at the connected points, terminates at the border of 
$B$ at $t=0$, and starts at the border of $A$ at $t=i\tau_{0}$. The time $\tau_{0}$ is defined by the total energy $E_{0}$.
  
The above general scheme is applicable to a calculation of tunneling rate regardless of dimensionality of the problem 
(we do not consider a possibility of caustics between $A$ and $B$). Nevertheless, when the dimensionality is bigger 
than one, another way may exist to tunnel through a static potential barrier. This way differs from the normal scheme and 
may result in an enhanced tunneling rate. Any tunneling process obeys quantum mechanical rules: a tunneling particle at each 
point emits partial de Broglie waves which interfere and produce a final outgoing wave. This wave is exponentially small for 
semiclassical one-dimensional barriers. In two- or multi-dimensional case the tunneling motion can be influenced by other 
(non-tunneling) motions leading to different conditions of interference. As a result, the outgoing wave may be proportional 
to the enhanced exponent, which is much bigger than normal one or even of the order of unity. The phenomenon of enhanced 
tunneling is considered below.
\subsection{Tunneling in a quantum wire}
\label{subsec:wire}
\begin{figure}[p]
\begin{center}
\vspace{0.6cm}
\leavevmode
\epsfxsize=\hsize
\epsfxsize=16cm
\epsfbox{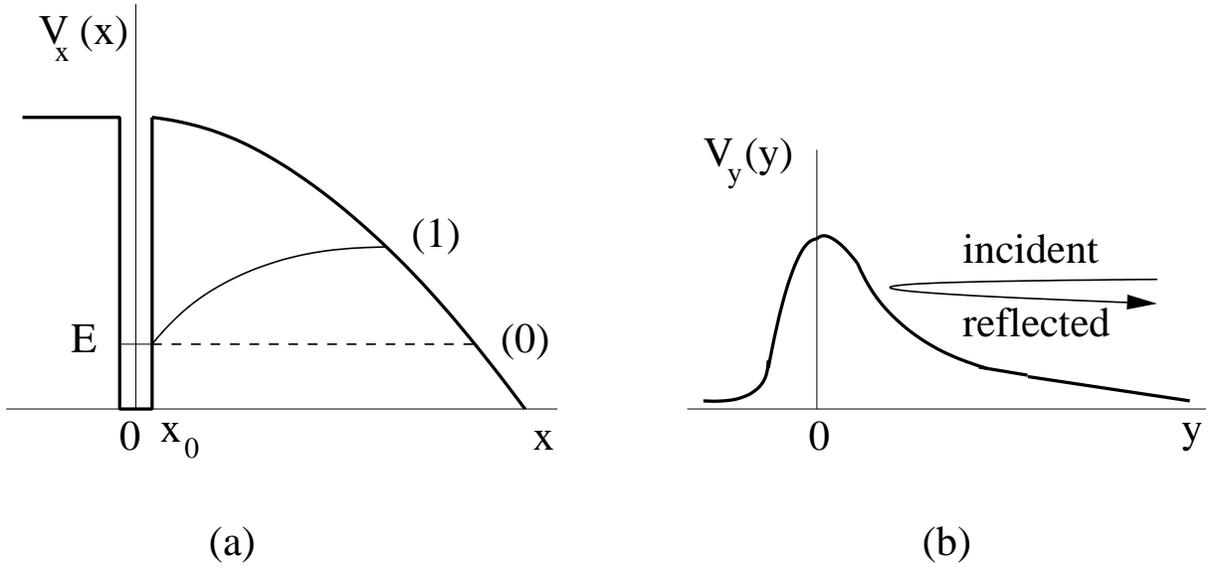}
\vspace{1cm}
\caption{The potential energy of a particle in a quantum wire $V_{x}(x)+V_{y}(y)$ corresponds to the quantized motion in the
radial $x$-direction (a) and an almost classical motion in the axis $y$-direction (b).}
\label{fig8}
\end{center}
\end{figure}

\begin{figure}[p]
\begin{center}
\vspace{1.5cm}
\leavevmode
\epsfxsize=\hsize
\epsfxsize=10cm
\epsfbox{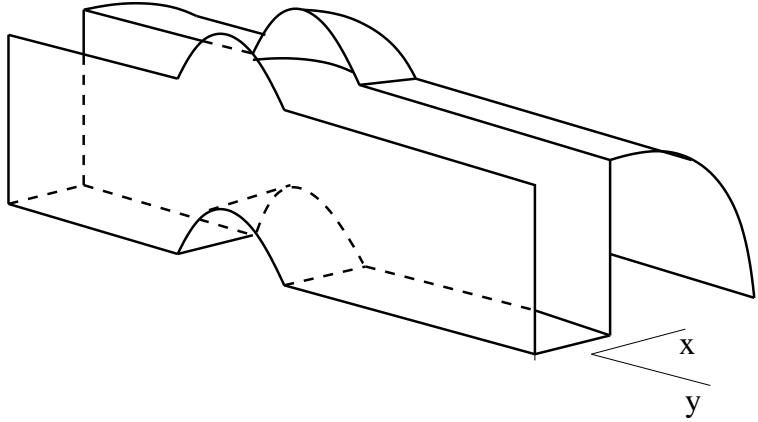}
\vspace{0.7cm}
\caption{The potential energy $V(x,y)$ of a particle in the quantum wire in two dimensions.}
\label{fig9}
\end{center}
\end{figure}

The mechanism of enhanced tunneling through a static barrier can be shown in the case of a long and narrow quantum wire 
when the radial motion ($x$-direction) is strongly quantized but the motion along the axis ($y$-direction) is almost 
classical. We do nod discuss a detailed link to real quantum wires since the goal is solely to demonstrate a new mechanism 
of tunneling through static barriers. To model this situation, suppose the particle to move in the two-dimensional potential 
$V_{x}(x)+V_{y}(y)$ plotted in Fig.~\ref{fig8}. The particle can tunnel from the energy level $E$ in the radial direction. In 
two dimensions the potential energy is shown in Fig.~\ref{fig9}. The particle tunnel from the region of small $x$ to the 
region of big $x$. When the total potential energy is simply a sum of $x$-dependent and $y$-dependent potentials, variables 
$x$ and $y$ are separated and tunneling along $x$-direction occurs as in a one-dimensional case.  
\subsection{Mapping of the stationary problem onto dynamical pulses}
\label{subsec:mapping}
If to add some ``interaction'' potential $V_{int}(x,y)$ to the system, the total potential
\begin{equation}
\label{34}
V(x,y)=V_{x}(x)+V_{y}(y)+V_{int}(x,y)
\end{equation}
is not reduced to a sum of $x$- and $y$-term and motions in the $x$- and $y$-direction are no more independent. The both
barriers in Fig.~\ref{8} are weakly penetrated and, in the first approximation, the particle occupies the region of
$x\sim x_{0}$ and positive $y$ in Fig.~\ref{9} with incident and reflected fluxes along $y$-direction as shown in 
Fig.~\ref{fig8}(b). This is equivalent to the problem of two one-dimensional particles with the interaction $V_{int}(x,y)$
when one particle in a well is acted by a steady flux of incident particles. This is analogous to the situation in nuclear
physics when an incident steady flux of protons acts on the alpha particle occupied some energy level inside the nuclear 
potential well \cite{WEISSKOPF,BOHR}. This nuclear problem of excitation of alpha particle can be considered by two methods: 
(i) by a solution of the Schr\"{o}dinger equation for alpha particle and proton in the static potential or (ii) by a solution 
of the Schr\"{o}dinger equation for alpha particle in a nonstationary potential created by the classically moved proton which 
reflects by nuclear Coulomb forces. The both ways lead to the same small exponent in the semiclassical probability. The second 
method is more convenient since the excitation of a higher level of alpha particle can be considered as one driven by a 
nonstationary force, acting during a finite time \cite{LANDAU}. So, with exponential accuracy, the steady problem of
decay of a metastable state under an incident flux can be mapped onto dynamical pulses. 

Under conditions, when the motion in the $y$-direction is weakly violated by the motion in the $x$-direction, the total
process can be considered as one-dimensional tunneling in the nonstationary potential $V_{x}(x)+V_{int}[x,y(t)]$, where 
$y(t)$ is a classical trajectory in the static potential $V_{y}(y)$, shown in Fig.~\ref{8}(b). According to this, besides 
normal tunneling ``0'' in Fig.~\ref{8}(a), corresponding to absence of the pulse, there is also the process ``1'' related 
(in the language of dynamical pulses) to the under-barrier photon-assisted tunneling which is similar to Fig.~\ref{fig4}. 
Since a photon-assistance increases tunneling probability, the process ``1'' relates to the phenomenon of enhanced
tunneling. In the language of static potential, one can say that tunneling along the radius is assisted by the motion along 
the axis. 

This mapping is mentioned only for interpretation of enhanced tunneling ``1'' by means the language of dynamical pulses.
In \ref{subsec:newtunnel} enhanced tunneling is considered on the basis of a static barrier.
\subsection{Probability of the enhanced tunneling}
\label{subsec:newtunnel}
\begin{figure}[p]
\begin{center}
\leavevmode
\epsfxsize=\hsize
\epsfxsize=8cm
\epsfbox{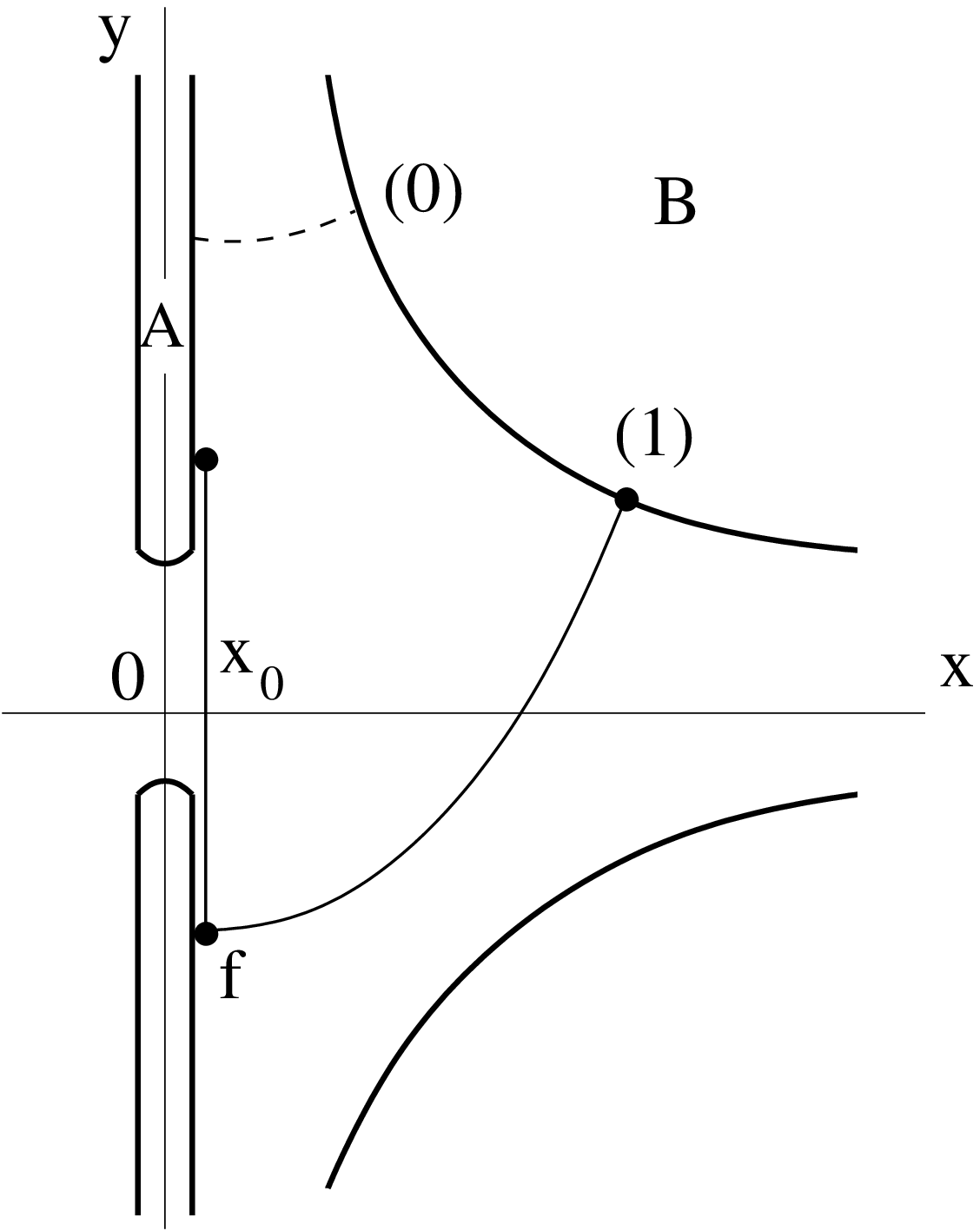}
\vspace{1cm}
\caption{Thick curves are determined by the condition $V(x,y)=E_{0}$. Normal tunneling ``0'' simply  relates to the 
classical trajectory denoted by the dashed curve. Enhanced tunneling ``1'' is more complicated and consists of two 
steps: $A\rightarrow f$ and the classical trajectory $f\rightarrow B$. The vertical line goes from $y=y_{1}$ down to 
$y=y_{0}$.}
\label{fig10}
\end{center}
\end{figure}

The condition $V(x,y)=E_{0}$ gives the curve shown in Fig.~\ref{fig10}. To calculate a probability of enhanced tunneling 
``1'', one should account the fact that in semiclassical limit the phase $S/\hbar$ of the wave function 
$\psi (x,y)\sim\exp[iS(x,y)/\hbar]$ is big. The function $S(x,y)$ can be found from Scr\"{o}dinger equation as the main term 
in the expansion of the type (\ref{14}). For normal tunneling ``0'', $S(x,y)$ coincides with the classical action, varies 
smoothly between the regions $A$ and $B$, and its extreme value relates to the classical trajectory denoted by the dashed curve 
in Fig.~\ref{fig10}. For enhanced tunneling ``1'' the behavior of $S(x,y)$ is much more complicated, since there are reconnection 
among different branches of $S$. On each branch the function $S(x,y)$ coincides with some classical action according to 
findings of Refs.~\cite{IVLEV1,IVLEV2}. 

Fortunately, to find a tunneling probability it is not necessary to know $\psi (x,y)$ at all $x$ and $y$. It is sufficient to 
connect phases of the wave function between certain points at the borders of $A$ and $B$ using some formal method. Such formal 
method for the process ``1'' (enhanced tunneling) consists of two steps: a connection $A\rightarrow f$ and a connection 
$f\rightarrow B$ (Fig.~\ref{fig10}). For the first step, one can consider the vertical line at $x=x_{0}$ connecting the border 
of $A$ and some point $f$ in Fig.~\ref{fig10}, where $\psi\sim\exp(i\sigma/\hbar)$ is expressed through the certain branch 
$\sigma$ of the classical action, which satisfies the equation of Hamilton-Jacobi
\begin{equation}
\label{35}
\frac{1}{2m}\left(\frac{\partial\sigma}{\partial x}\right)^{2}+\frac{1}{2m}\left(\frac{\partial\sigma}{\partial y}\right)^{2}
+V(x,y)=E_{0}
\end{equation}
At $x>x_{0}$  the potentials $V_{x}(x)$, $V_{y}(y)$, and $V_{int}(x,y)$ are smooth. Since the potential $V_{x}(x)$ is 
abrupt at $x=x_{0}$, existence of the quantum level $E$ in the well results in the following semiclassical condition inside 
the barrier in Fig.~\ref{8}(a), when $x$ tends to $x_{0}$ from the right,
\begin{equation}
\label{36}
\frac{1}{2m}\left(\frac{\partial\sigma}{\partial x}\right)^{2}\bigg |_{x_{0}}+V_{x}(x_{0})=E
\end{equation}
The total energy is $E_{0}=E+\varepsilon$, where $\varepsilon$ is the energy of the classical motion along the well. The 
$y$-dependence of the action at $x=x_{0}$ can be easily found from Eqs.~(\ref{35}) and (\ref{36}) in the WKB form of the 
one-dimensional motion along the well. The phase at the point $f$ (relative to $A$) in Fig.~\ref{fig10} is
\begin{equation}
\label{37}
\frac{\sigma}{\hbar}=-\frac{i}{\hbar}\int^{y_{1}}_{y_{0}}dy\sqrt{2m\left[V_{y}(y)+V_{int}(x_{0},y)-\varepsilon\right]}
\end{equation}
where $y_{0}$ and $y_{1}$ are determined by zero of the square root in Eq.~(\ref{37}) at $f$ and $A$ respectively. The 
position of the point $f$, defined by the conditions $x=x_{0}$ and $y=y_{0}$, is not flexible and depends only on the 
energy $\varepsilon$, since the quantum level $E$ is strictly determined by the potential in Fig.~\ref{8}(a). The same 
relates to the phase $\sigma/\hbar$ at the point $f$ (relative to $A$). The value of $\sigma$, given by Eq.~(\ref{37}), 
is extreme with respect to variation of positions of the ends of the line $x=x_{0}$ in Fig.~\ref{fig10}.

As the second step, the point $f$ can be connected to the region $B$ by the classical trajectory which provides an extreme 
of the phase of $B$ (relative to $f$) and is based on the expression $A_{0}(f\rightarrow B)$, as Eq.~(\ref{32}), but the 
trajectory connects now the regions $f$ and $B$. The rate of enhanced tunneling ``1'' $\exp(-A_{1})$ is given by the 
total phase difference and reads
\begin{equation}
\label{38}
A_{1}=A_{0}(f\rightarrow B)-\frac{2i\sigma}{\hbar}
\end{equation}
where $\sigma$ is defined by Eq.~(\ref{37}). As soon as normal tunneling is described solely by a classical trajectory 
(\ref{32})-(\ref{33}), enhanced tunneling, according to Eq.~(\ref{38}), requires more complicated description. The total
tunneling probability should be estimated as
\begin{equation}
\label{38a}
W\sim\max\left\{\exp(-A_{0});\hspace{0.1cm}\exp(-A_{1})\right\}
\end{equation}
To calculate $A_{0}$ (normal) and $A_{1}$ (enhanced) one should specify the potential energy in the form (\ref{34}) and 
use Eqs.~(\ref{32}), (\ref{33}), (\ref{37}), and (\ref{38}). 

The second term in Eq.~(\ref{38}), which is real and negative, works versus the big first term representing the double tunneling 
through the barriers in Figs.~\ref{8}(a) and \ref{8}(b). The second term in Eq.~(\ref{38}) can play a crucial role in reduction 
of $A_{1}$. Enhanced tunneling ``1'' may be interpreted as a cooperative motion along a trajectory in the $x$-direction and along 
a bounce (forth and back) in the $y$-direction. One should emphasize again, that the above semiclassical method gives only the 
formal calculation of the tunneling probability with exponential accuracy, but it does not relate to a real behavior of the wave 
function $\psi (x,y)$. For example, the formal density at the point $f$ can be big, the real wave function drops exponentially in 
that region. 
\section{DISCUSSION AND CONCLUSIONS}
There are two issues in this paper.

The first one is that the problem of inter-band tunneling in a semiconductor (Zener breakdown) in a nonstationary and 
homogeneous electric field is solved exactly. On the basis of the exact analytical solution the formation of an approximation 
of classical trajectories is studied. This semiclassical formalism is very effective in tunneling under nonstationary 
conditions since it allows to reduce the problem to a solution of the Newton equation in complex time. The semiclassical 
approach, obtained from the exact solution, is of the same type as in other problems of nonstationary tunneling which do 
not allow exact solutions, for example, decay of a metastable state, penetration through a potential barrier, and 
over-barrier reflection.

The second issue relates to different types of tunneling through a static potential barrier if dimensionality of the problem
is bigger than one. A multi-dimensional case has an essential feature which differs it from one-dimensional problem. Namely,
a tunneling coordinate $(x)$ can be influenced by non-tunneling one $(y)$ resulting in a different condition of interference 
of emitted partial waves during tunneling through the potential $V(x,y)$. A multi-dimensional tunneling can be interpreted as 
one-dimensional but in the nonstationary potential $V[x,y(t)]$, where $y(t)$ is the classical non-tunneling motion. This mapping 
onto dynamical problem enables to establish different types of tunneling. 

As well known, there is {\it normal} tunneling through a multi-dimensional barrier related simply to a classical 
trajectory in imaginary time and connected two classically allowed regions. This mechanism was always employed by authors 
in consideration of multi-dimensional tunneling. Normal tunneling corresponds to the case when the tunneling particle does
not absorb quanta of the nonstationary field. 

There is also a process when particle absorbs quanta (photons) from the nonstationary potential $V[x,y(t)]$. This is a 
photon-assisted tunneling with an enhanced probability. In the language of the static potential barrier, this process
corresponds to {\it enhanced} tunneling. This is a new mechanism of quantum tunneling through a static multi-dimensional
barrier. Depending on shape, sign, and value of the potential $V(x,y)$, two types of enhanced tunneling are possible: (i) 
with absorption of energy by the tunneling motion (positive assistance) or (ii) with emisssion of energy (negative assistance). 
The positive assistance was considered in Ref.~\cite{IVLEV1} and in this paper for Zener breakdown. The negative assistance was 
considered in Ref.~\cite{IVLEV2}. It is remarkable that the method of calculation of probability of enhanced tunneling, proposed 
in this paper, is applicable equally to both positive and negative assistance. 

As it was found in Ref.~\cite{IVLEV2} for tunneling through a nonstationary barrier, the enhanced tunneling probability may be 
not exponentially small under conditions of negative assistance. This phenomenon is called Euclidean resonance which results
from influence of nonstationary field on interference of emitted partial waves during tunneling. Euclidean resonance can also 
take place in tunneling through a static non-one-dimensional barrier. To get this phenomenon, one should adapt barrier 
parameters and the particle energy to reach the condition $A_{1}=0$ in Eq.~(\ref{38}), as in the case of a nonstationary barrier
\cite{IVLEV2}. The above theory is valid if $\exp(-A_{1})\ll 1$ and when this condition breaks down one should use a formalism 
generic with the multi-instanton approach. Nevertheless, the presented theory, as any other approach, can be approximately 
extended up to the limit of its validity $\exp(-A_{1})\sim 1$ wich relates to Euclidean resonance. 

Enhanced tunneling may result in a dramatic increase of a tunneling rate. For example, for an almost classical barrier, 
according to the normal mechanism, the tunneling probability is calculated to be $10^{-100}$, but enhanced tunneling, under 
conditions of Euclidean resonance, results in the probability of $10^{-2}$. Euclidean resonance is a phase phenomenon and any 
effect of dephasing (friction, for example) may destroy it. To avoid this destruction, the attenuation time due to friction 
should be bigger compared to the under-barrier time. 

In conclusion, a new mechanism of enhanced tunneling through static non-one-dimensional barriers is established in addition to 
well known normal tunneling. Under certain conditions on shape of an almost classical barrier and particle energy, the 
probability of quantum tunneling through it may be not exponentially small.
\acknowledgments
I am grateful to V. Gudkov for valuable discussions.

\newpage

\begin{references}

\bibitem{RABITZ}

W.S. Warren, H. Rabitz, and M. Dahlen, Science {\bf 259}, 1581 (1993).

\bibitem{SHI}

S. Shi and H. Rabitz, J. Chem. Phys. {\bf 92}, (1990).

\bibitem{JUDSON}

R.S. Judson and H. Rabitz, Phys. Rev. Lett. {\bf 68}, 1500 (1992).

\bibitem{KOHLER}

B. Kohler, J.L. Krause, F. Raksi, K.R. Wilson, V.V. Yakovlev, R.M. Whitnel, and \\
Y. Yan, Acc. Chem. Res. {\bf 28}, 133 (1995).

\bibitem{SCHUMACHER}

D.W. Schumacher, J.H. Hoogenraad, D. Pinkos, and P.H. Bucksbaum, Phys. Rev. A {\bf 52}, 4719 (1995).

\bibitem{KRAUSE}

J.L. Krause, D.H. Reitze, G.D. Sanders, A.V. Kuznetsov, and C.J. Stanton, Phys Rev. B {\bf 57}, 9024 (1998).

\bibitem{BERMAN}

T. Martin and G. Berman, Phys. Lett. A {\bf 196}, 65 (1994)

\bibitem{ATANASOV}

R. Atanasov, A. Hache, J.L.P. Hughes, H.M. van Driel, and J.E. Sipe, Phys. Rev. Lett. {\bf 76}, 1703 (1996).

\bibitem{LANDAU}

L.D. Landau and E.M. Lifshitz, {\it Quantum Mechanics} (Pergamon, New York, 1977).

\bibitem{POKR}

V.L. Pokrovskii, F.R. Ulinich, and S.K. Savvinykh, Zh. \'{E}ksp. Teor. Fiz. {\bf 34}, 1629 (1958) 
[Sov. Phys. JETP {\bf 34}, 1119 (1958)].

\bibitem{COLEMAN}

C.G. Callan and S. Coleman, Phys. Rev. D {\bf 16}, 1762 (1977).

\bibitem{KELDYSH}

L.V. Keldysh, Zh. \'{E}ksp. Teor. Fiz. {\bf 47}, 1945 (1964) [Sov. Phys. JETP {\bf 20}, 1307 (1965)].

\bibitem{PERELOMOV}

V.S. Popov, V.P. Kuznetsov, and A.M. Perelomov, Zh. \'{E}ksp. Teor. Fiz. {\bf 53}, 331 (1967) 
[Sov. Phys. JETP {\bf 26}, 222 (1968)].

\bibitem{MELN1}

B.I. Ivlev and V.I. Melnikov, Pis'ma Zh. \'{E}ksp. Teor. Fiz. {\bf 41}, 116 (1985) [JETP Lett. {\bf 41}, 142 (1985)]

\bibitem{MELN2}

B.I. Ivlev and V.I. Melnikov, Phys. Rev. Lett. {\bf 55}, 1614 (1985)

\bibitem{MELN3}

B.I. Ivlev and V.I. Melnikov, Zh. \'{E}ksp. Teor. Fiz. {\bf 90}, 2208 (1986) [Sov. Phys. JETP {\bf 63}, 1295 (1986)]. 

\bibitem{MELN4}

B.I. Ivlev and V.I. Melnikov, Phys. Rev. B {\bf 36}, 6889 (1987)

\bibitem{MELN5}

B.I. Ivlev and V.I. Melnikov, in {\it Quantum Tunneling in Condensed Media}, edited by\\ 
A. Leggett and Yu. Kagan (North-Holland, Amsterdam, 1992).

\bibitem{MILLER}

W.H. Miller, Adv. Chem. Phys. {\bf 25}, 68 (1974).

\bibitem{KESHA}

S. Keshavamurthy and W.H. Miller, Chem. Phys. Lett. {\bf 218}, 189 (1994).

\bibitem{DEFENDI}

A. Defendi and M. Roncadelli, J. Phys. A {\bf 28}, L515 (1995).

\bibitem{MAITRA}

N.T. Maitra and E.J. Heller, Phys. Rev. Letter. {\bf 78}, 3035 (1997).

\bibitem{ANKERHOLD}

J. Ankerhold and H. Grabert, Europhys. Lett. {\bf 47}, 285 (1999).

\bibitem{CUNIBERTI}

G. Cuinberty, A. Fechner, M. Sassetti, and B. Kramer, Europhys. Lett. {\bf 48}, 66 (1999).

\bibitem{IVLEV1}

B.I. Ivlev, Phys. Rev. A {\bf 62}, 062102 (2000).

\bibitem{IVLEV2}

B.I. Ivlev, Phys. Rev. A {\bf 66}, 012102 (2002).

\bibitem{HEADING}

J. Heading, in {\it An introduction to Phase-Integral Method}, edited by Methuen\\ 
(Wiley, New York, 1962)

\bibitem{ZIMAN}

J.M. Ziman, {\it Principles of the Theory of Solids} (Cambridge University Press, 1964) 

\bibitem{FEYNMAN}

R.P. Feynman and A.R. Hibbs, {\it Quantum Mechanics and Path Integrals} (McGrow-Hill, New York, 1965)

\bibitem{KAGAN}

I.M. Lifshitz and Yu. Kagan, Zh. \'{E}ksp. Teor. Fiz. {\bf 62}, 1 (1972) [Sov. Phys. JETP {\bf 35}, 206 (1972)] 

\bibitem{POKR}

B.V. Petukhov and V.L. Pokrovsky, Zh. \'{E}ksp. Teor. Fiz. {\bf 63}, 634 (1972) [Sov. Phys. JETP {\bf 36}, 336 (1973)]

\bibitem{OKUN}

M.B. Voloshin, I.Yu. Kobzarev, and L.B. Okun, Yad. Phys. {\bf 20}, 1229 (1974) [Sov. J. Nucl. Phys. {\bf 20}, 644 (1975)]

\bibitem{STONE}

M. Stone, Phys. Rev. D {\bf 14}, 3568 (1976).

\bibitem{COLEMAN}

C.G. Callan and S. Coleman, Phys. Rev. D {\bf 16}, 1762 (1977).

\bibitem{SCHMIDT}

A. Schmidt, in {\it Quantum Tunneling in Condensed Media}, edited by A. Leggett and Yu. Kagan 
(North-Holland, Amsterdam, 1992).

\bibitem{MELN6}

B.I. Ivlev and V.I. Melnikov, Phys. Rev. B {\bf 36}, 6889 (1987).

\bibitem{OVCH}

B.I. Ivlev and Yu.N. Ovchinnikov, Zh. \'{E}ksp. Teor. Fiz. {\bf 93}, 668 (1987) [Sov. Phys. JETP {\bf 66}, 378 (1987)] 

\bibitem{WEISSKOPF}

J.M. Blatt and V.F. Weisskopf, {\it Theoretical Nuclear Physics} (Springer-Verlag,\\ 
New York, 1979)

\bibitem{BOHR}

A. Bohr and B.R. Mottelson, {\it Nuclear Structure} (W.A. Benjamin, New York,\\ 
Amsterdam, 1969)


\end{references}
\end{document}